\documentclass[preprint,aps]{revtex4}
\usepackage{amssymb,amsmath}
\usepackage{graphicx,float}

\begin{document}
\title{ \Large \bf Exploring Complex Graphs by Random Walks \footnote[3]{based on talk presented at Seventh Granada Lectures {\bf Modeling Complex Systems},
Granada, Spain 2002; Appeared (without abstract!) in {\bf Modeling Complex Systems}, AIP Conference Proceedings 661, p. 24, (2003), Editors: P.L. Garrido and J. Marro.}}

\author{Bosiljka Tadi\'c$^*$ }
\affiliation{Jo\v zef Stefan Institute, 1001 Ljubljana, Slovenia, $^*$e-mail: Bosiljka.Tadic@ijs.si}


\begin{abstract}
We present an algorithm \cite{BT} to grow a graph with scale-free structure
of {\it in-} and {\it out-links} and variable wiring diagram in the 
class of the world-wide Web. We then explore the graph by  
intentional random walks using local next-near-neighbor search algorithm 
to navigate through the graph. The topological properties such as 
betweenness are determined by an ensemble of independent walkers and 
efficiency of the search is compared on three different graph topologies.
In addition we simulate interacting random walks which are created by 
given rate and navigated in parallel, representing transport with queueing of 
information packets on the graph. 
\end{abstract}
\maketitle

\section{Introduction}
We are living in the world of networks, majority of which are not static but 
evolve in time \cite{recent_review}. The evolution of network is governed 
by microscopic rules that in course of time lead to the emergence of complex 
structures through the mechanism of self-organization and dynamical 
constraints.
The emergent scale-free structure of links in  technological
networks the Internet and  the Web and in molecular and gene regulatory 
networks is of particular importance for their functional stability. 
The networks are
adequately represented by graphs with a given structure of links. The obvious
question is then how the networks function, making the necessity of modeling
dynamic processes on these graphs. 
The dynamics of random walks on graphs represents a powerful tool, in which
we can explore connection between 
the graphs topology  and functional properties of the network.  

To explore  numerically graphs representing complex evolving networks 
by random walks, two types of algorithms are necessary: First, an algorithm to 
grow the graph with a given structure of links; Second, an algorithm that
governs a  random walk on that graph. Complexity of link structure in 
scale-free graphs such as the Web graph
suggests various possibilities for local navigation. For instance, apart 
from a naive move strategy, the random walker may learn from nodes linking 
preferences to get quickly to a well connected node
\cite{adaptive}.  For {\it intentional} random walks directed to a given
destination on the graph, we implement {\it local search algorithm} that
monitors near- and next-near neighbors in search for the destination node.

%
%
%
%

\section{Graph growth algorithms: the Web graph}
We present an algorithm \cite{BT} to grow a graph with scale-free 
structure and flexible wiring diagram in the class of the world-wide Web.
Objectives are to grow a graph that has statistically the same properties
as measured in the real Web: scale-free degree distributions both for 
in- and out-links (exponents $\tau_{in}\approx2.2$
and $\tau_{out}\approx 2.6$); occurrence of a giant component and the 
distribution of clusters with the exponent $\tau_s\approx
2.5$. As demonstrated in Ref.\ \cite{BT} a minimal set of  microscopic rules
necessary to reproduce such graphs  include 
{\it growth, attachment}, and {\it rewiring}. 
Time is measured by addition of a node, which  attempts to link with
probability ${\tilde{\alpha }}$ to a node $k$. Else, with probability
$1-{\tilde{\alpha }}$ a preexisting node $n$ rewires or adds a new out-link 
directed to $k$. Nodes $k$ and $n$ are selected with  probabilities 
$p_{in}\equiv p_{in}(k,t)$, $p_{out}\equiv p_{out}(n,t)$
\begin{equation}
p_{in}=  (M\alpha + q_{in}(k,t))/(1+\alpha )Mt \   ;   
p_{out}= (M\alpha + q_{out}(n,t))/(1+\alpha )Mt \ ,
\label{Pt}
\end{equation}
which depend on current number of respective links $q_{in}(k,t)$ and  
$q_{out}(n,t)$. $M$ is average number of links  per time step 
(see \cite{BT,BT_ccp} for more details). The graph flexibility, which is
measured by the degree of 
rewiring $(1-{\tilde{\alpha }})/{\tilde{\alpha }}$, is 
essential both for the appearance of the scale-free structure
of {\it out-links} and for occurrence of closed cycles, 
which affect the dynamic processes on the graph. 

By solving the corresponding rate equations we find that 
the local connectivities $<q_{in}(s,t)>$ and $<q_{out}(s,t)>$ at 
a node added at time $s$ increase with time $t$ as 
\begin{equation}
q_\kappa (s,t) = A_\kappa \left[\left(t/s\right)^{\gamma _\kappa} 
-B_\kappa \right] \ .
\label{qst}
\end{equation}
with $\kappa = in$ and $out$, and $\gamma _{in} = 1/(1+\alpha )$ and 
$\gamma _{out} = (1-{\tilde{\alpha }})/(1+\alpha )$.  We use the original
one-parameter model introduced in \cite{BT} with ${\tilde{\alpha }} = 
\alpha =0.25$ and $M=1$. 
When ${\tilde{\alpha }} =1$ the emergent structure is tree like with 
one out-link per node.

\begin{figure}[htb]
\begin{center}
\begin{tabular}{cc} 
\includegraphics[width=7.5cm]{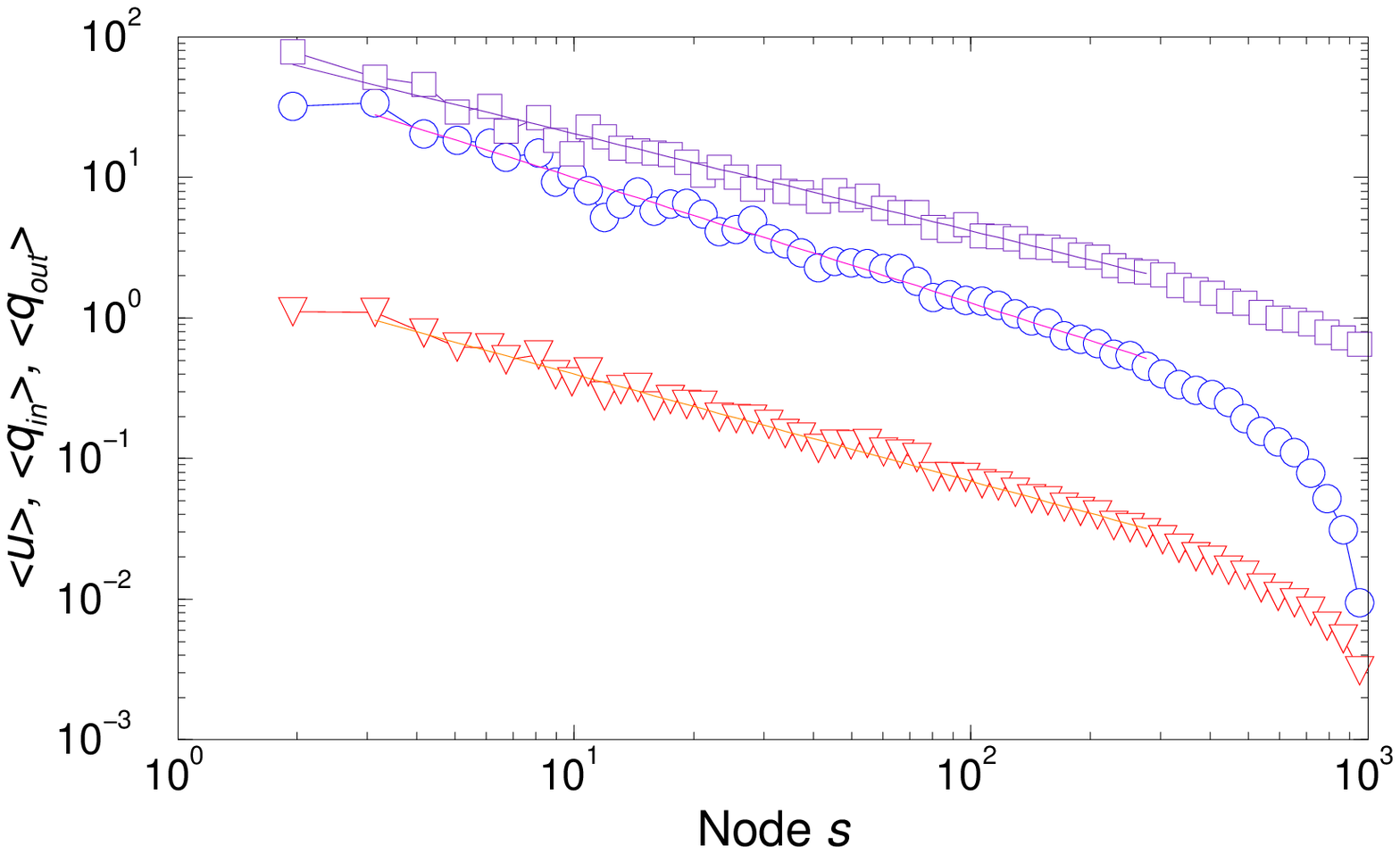}&
\includegraphics[width=7.5cm]{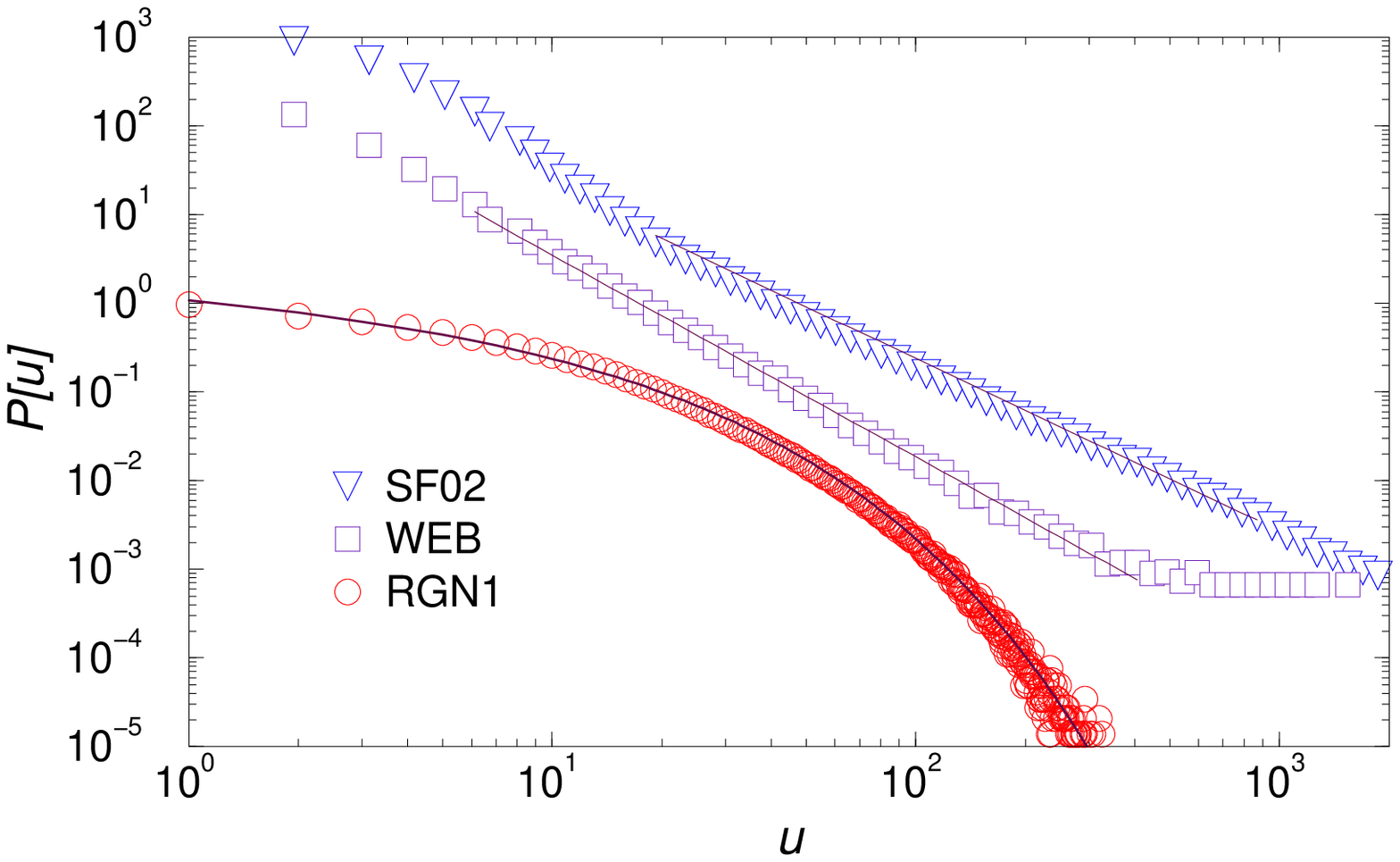} \\
\end{tabular}
\end{center}
\caption{Left panel: Average connectivity of  node $s$ for (top to bottom)
out-links, in-links and number of visits $<u>$ for sequential walks
in the Web graph.
Right panel: Probability distribution $P(u)$ of visits $u$ to a given 
node vs $u$ in the Web graph, tree graph with same in-link distribution
 and for randomly grown graph. Curves shifted vertically. All data log-binned.}
\label{figgraph}
\end{figure}

In Fig.\ 1a we show
simulated local connectivities for $t=N=10^4$ nodes
in agreement with Eq.\ (\ref{qst}). This implies the power-law behavior 
of emergent degree distributions with exponent $\tau_\kappa =
1/\gamma _\kappa +1$, in agreement with  simulations in \cite{BT,BT_ccp}.
\section{Random Walks on scale-free graphs}

We now explore the graph with an ensemble of {\it intentional} random walks.
 To stress importance of the graph topology, we compare results
obtained on the Web graph, the tree graph with same in-link structure and on
randomly grown graph.   We first consider dynamics of an ensemble of
independent random walkers, that can be implemented sequentially.
 We then introduce  some details of the dynamics of interacting 
random walks which are navigated in parallel while moving towards their 
selected destinations, leading to queueing processes on the Web graph.

\subsection*{Sequential Walks and Graph Topology}

For each random walk we select a start and destination node at random
(see \cite{TP,book} for electrical networks). 
Then the walk is navigated by next-near-neighbor ($nnn$-) algorithm to 
search for 
its destination. In Fig.\ 1a also is shown the average number of visits
$<u>$ at node $s$ for the walker using both in- and out-links 
equally. As expected, the number of visits to a node are 
determined by its local connectivities. This is further illustrated in
 Fig.\ 1b where the distribution of number of visits $P(u)$ is shown for
three different graph topologies. The distribution falls of faster in the
case of Web (slope $\tau_u =2.275\pm 0.015$) compared to the tree graph
($\tau _u =1.94\pm 0.018$), suggesting that the presence of closed cycles 
on the Web graph increases its searcability.

\begin{figure}[htb]
\begin{center}
\begin{tabular}{cc} 
\includegraphics[width=8cm]{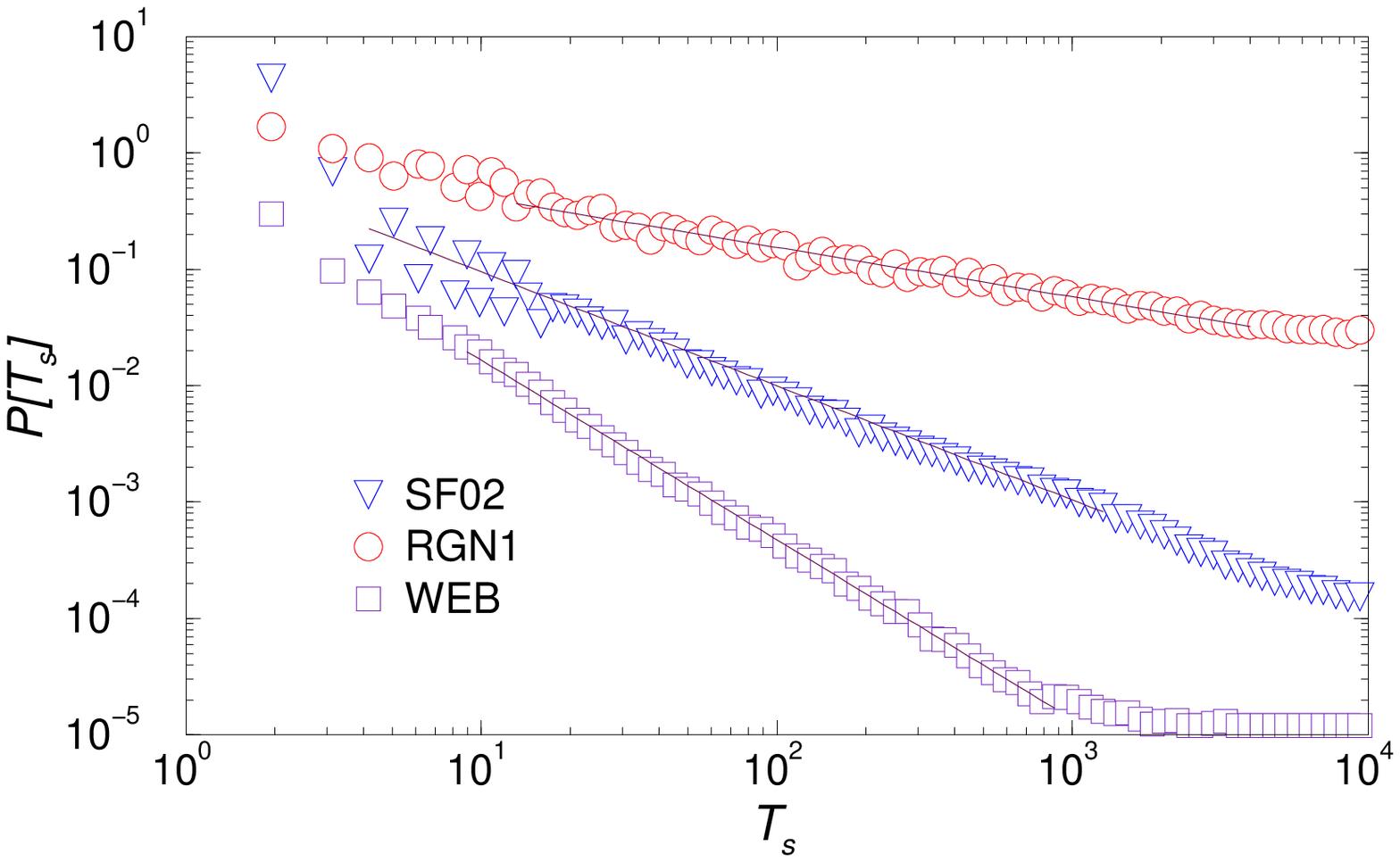}&
\includegraphics[width=8cm]{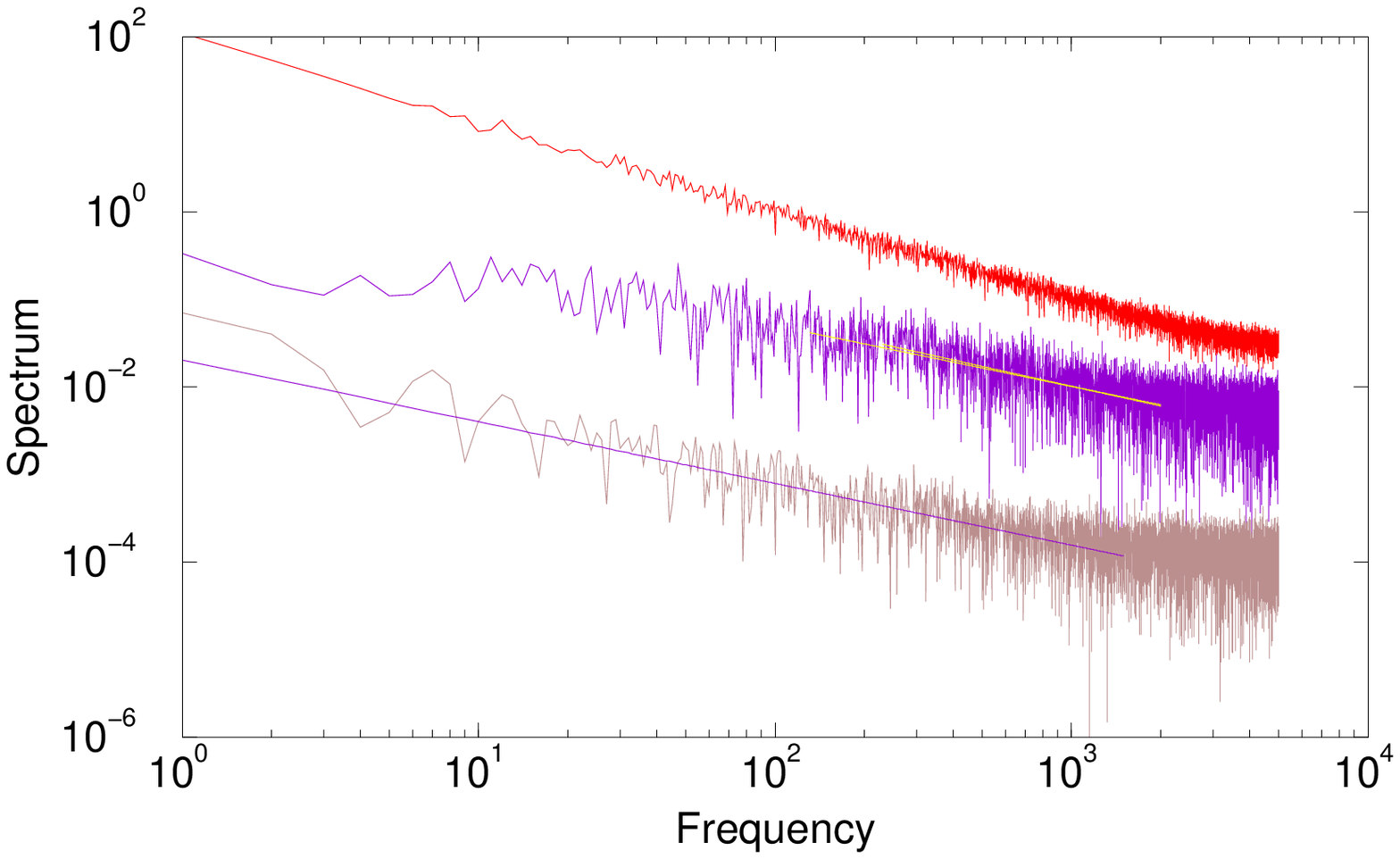} \\
\end{tabular}
\end{center}
\caption{Left panel:  Probability distribution of search times $P(T_s)$ vs  $T_s$ for three graph topologies as in Fig.\ 1b. All data are logarithmically binned. Right  panel: Spectrum of the density, number of active nodes, 
and number of packets in the traffic (bottom to top).}
\label{figgraph}
\end{figure}

The higher efficiency of the $nnn$-search on the Web compared to tree 
graph is further expressed in the distributions of  duration of search 
on Fig.\ 2a, where the respective slopes are $\tau_{ts}=1.541\pm 0.008$ and 
$\tau_{ts}=0.98\pm 0.016$ .  Intuitively, this can be understood
as the $nnn$-search algorithm often directs walks through the hub node,
to which a majority of in-links point. In the Web graph, the same
algorithm can use two types of well connected nodes---hub and authority nodes.
It is interesting to note that a broad distribution of search time is also 
found  on randomly grown graph (the exponent  $\tau_{ts}=0.425\pm 0.008$), 
whereas the corresponding distribution of visits is 
a stretch-exponential function (cf. Fig.\ 1b).

\subsection*{Parallel Walks: Transport on Graphs}
To model simultaneous random walks, as for instance information
packet transport on the Internet, in addition to graph topology and search 
algorithm as above, we need additional parameters to specify packet creation 
rate $R$ and queueing discipline, as LIFO \cite{TR}. In addition, we assume
finite maximal queue length $H$, which then  modifies search  when a local 
queue is at its maximum \cite{TR}.  Technically, the object-oriented
programming is necessary.
As a consequence of simultaneous intentional walks, we have  new phenomena
related to {\it dynamically interacting queues on the graph}. 
At low creation rate $R\to 0$
the statistics of independent walks as above applies. For $R > 0$, the time 
that a walker spends on the graph, which is interesting for costs planning,
  consists of search time and time that it spends trapped in queues along the 
way. For large $R$ we have dense traffic, in which searchability is limited,
and eventually a transition to jammed state occurs. 
The dynamics of queueing depends on the graph topology and the parameters 
$R$ and $H$. Few details are shown in Fig.\ 2b 
for our  scale-free graph with cycles. For large enough creation rate 
the density of packets arriving at a hub node in a single time step
 exhibits long-range temporal correlations. Similar correlations are found 
in the number of simultaneously moving packets (activity) and in the number 
of present packets
(moving and queueing) at each time step. An analysis of transit time statistics and queue properties  for a scale-free tree graph can be found in \cite{TR}.

Presence of closed cycles enhances searchability of flexible 
scale-free graphs. The efficiency of $nnn$-search algorithm on these graphs 
is related to occurrence of hub and authority nodes.
Sequential walks discover  directly the graph topology, which together  
with parameters $R,H$ affects  transport and  queueing on the graph.

\end{document}